\documentclass{article}
\PassOptionsToPackage{numbers,compress}{natbib}

\usepackage[preprint]{neurips_2026}

\usepackage[utf8]{inputenc}
\usepackage[T1]{fontenc}
\usepackage{hyperref}
\usepackage{url}
\usepackage{booktabs}
\usepackage{amsfonts}
\usepackage{nicefrac}
\usepackage{microtype}
\usepackage{xcolor}
\usepackage{amssymb}
\usepackage{amsmath}
\usepackage{graphicx}
\usepackage{float}
\usepackage{listings}
\usepackage{makecell}

\hypersetup{
    hidelinks
}
\title{Predictive Maps of Multi-Agent Reasoning:\\
A Successor-Representation Spectrum for LLM Communication Topologies}

\author{%
Ethan Parks \\
\texttt{edparks@arizona.edu} \\
University of Arizona \\
\And
Dalal Alharthi \\
\texttt{dalharthi@arizona.edu} \\
University of Arizona \\
}

\begin{document}
\maketitle

\begin{abstract}
Practitioners deploying multi-agent large language model (LLM) systems must currently choose between communication topologies (chain, star, mesh, and richer variants) without any pre-inference diagnostic for which topology will amplify drift, converge to consensus, or remain robust under perturbation. Existing evaluation answers these questions only post hoc and only for the task measured. We introduce a structural diagnostic for multi-agent LLM communication graphs based on the successor representation $M = (I - \gamma P)^{-1}$ of the row-stochastic communication operator, and we connect three of its spectral quantities, the spectral radius $\rho(M)$, the spectral gap $\Delta(M)$, and the condition number $\kappa(M)$, to three distinct failure modes. We derive closed-form spectra for the chain, star, and mesh under row-stochastic normalization, and validate the predictions on a 12-step structured state-tracking task with Qwen2.5-7B-Instruct over 100 independent trials. The condition number is a perfect rank-order predictor of empirical perturbation robustness ($r_s = 1.0$); the spectral gap partially predicts consensus dynamics ($r_s = 0.5$); and the spectral radius is perfectly \emph{inverted} with respect to cumulative error ($r_s = -1.0$). We trace this inversion to a regime in which linear spectra are blind to non-contracting bias drift, and we propose an affine-noise extension of the predictive map that recovers the empirical ordering. We read this as a first step toward representational, drift-aware structural diagnostics for multi-agent LLM systems, sitting alongside classical spectral and consensus theory.
\end{abstract}

\section{Introduction}
A recurring lesson across cognitive science, reinforcement learning, and systems neuroscience is that intelligent behavior is shaped as much by the \emph{structure of representation} as by the computations that operate over it \citep{stachenfeld2017hippocampus, momennejad2017successor}. An agent that carries a predictive map of its environment reasons differently from one that does not, even when both have access to the same local transitions. We bring this lens to a setting in which it has rarely been applied: multi-agent LLM systems, where a handful of language model instances exchange intermediate state through a fixed communication graph \citep{wei2023cot, besta2024graph, du2023debate, chan2023chateval, wu2023autogen, guo2024survey}. The graph is the structure. Our question is whether an appropriate representation of that structure, taken before any inference is run, can predict how the system will reason.

The practical motivation is concrete. A designer choosing between a 12-agent chain, a judge-and-leaves star, or a peer-deliberating mesh currently has no principled diagnostic for the question \emph{which topology amplifies drift, which converges to consensus, and which is brittle under perturbation}. Existing evaluation protocols answer these questions only after the pipeline has been executed, and only for the specific task measured \citep{liu2025agentbench, liang2023helm, srivastava2023beyond}. A structural diagnostic would close this loop, in the same spirit in which predictive maps close the loop between a transition structure and the behavior it supports.

We model a multi-agent LLM system as a directed graph whose nodes are agents and whose edges are information-passing channels, and we treat the row-normalized adjacency as a stochastic transition operator $P$. The successor representation $M = (I - \gamma P)^{-1}$, introduced in reinforcement learning by \citet{dayan1993sr} and subsequently shown in computational neuroscience to describe a predictive cognitive map in hippocampal circuits \citep{stachenfeld2017hippocampus, momennejad2017successor}, compresses all multi-step influence pathways into a single linear operator. Its eigenspectrum offers three scalar summaries with distinct mechanistic meaning: the spectral radius $\rho(M)$, the spectral gap $\Delta(M)$, and the condition number $\kappa(M)$. Each admits a hypothesis about a different failure mode of reasoning.

We study three canonical topologies, chain, star, and mesh, on a 12-step structured state-tracking task with Qwen2.5-7B-Instruct. The condition number rank-predicts empirical perturbation robustness perfectly; the spectral gap partially tracks consensus dynamics; and the spectral radius is \emph{anti-correlated} with cumulative error. The last finding is not a numerical artifact. It is a gap between the linear notion of stability that spectra encode and the sequential-drift notion of stability that LLM reasoning actually requires. The chain, which is maximally stable in the linear sense, licenses a monotone drift because each agent's small non-zero bias is handed forward without averaging. Star and mesh topologies include aggregation steps that function as implicit averaging and suppress the drift. We call this the \emph{stability paradox}, and we argue that it motivates a class of representational, drift-aware diagnostics that sit alongside classical spectral ones.

We are explicit that this is a case study. A single model family, one task family, three topologies, and 100 trials per condition do not establish a universal law, and a Spearman coefficient over three ranks carries, by construction, very little statistical weight. What we offer is a framework with derivable predictions and a first set of controlled experiments that instantiate it.

Our contributions are as follows. We formalize multi-agent LLM systems as row-stochastic communication graphs and introduce a successor-representation spectral diagnostic that compresses multi-step influence pathways into three scalar quantities. We derive closed-form spectra for the chain, star, and mesh topologies and pair each spectral quantity with a falsifiable empirical metric. We report a controlled study on a 12-step structured state-tracking task and find that the condition number perfectly rank-orders perturbation robustness, the spectral gap partially tracks consensus dynamics, and the spectral radius is exactly inverted with respect to cumulative error. We give a theoretical explanation of the inversion in terms of an affine-noise model in which each agent contributes iid noise that is reduced by aggregation but accumulated under sequential composition, and we show that the resulting drift-corrected gain rank-orders the empirical cumulative error correctly. We release task, code, and model configuration for reproduction. Taken together, these results convert topology selection from a post-hoc empirical question into a closed-form linear-algebraic one: given a candidate communication graph, three eigenvalue computations on $M = (I - \gamma P)^{-1}$ rank-order it against alternatives on drift, consensus, and adversarial sensitivity  before a single token is generated.

\section{Related Work}

\textbf{Multi-agent LLM systems.} A growing body of work composes
multiple LLM calls into explicit reasoning graphs: chain-of-thought
prompting established linear intermediate reasoning \citep{wei2023cot};
tree- and graph-of-thought generalized it to branching and merging
\citep{yao2023tot, besta2024graph}; debate, peer critique, and judge
aggregation introduced multi-agent deliberation
\citep{du2023debate, chan2023chateval, liang2024divergent, khan2024debating};
self-refinement internalized the deliberation loop
\citep{madaan2023selfrefine}; and engineering frameworks made
topologies first-class configuration choices
\citep{wu2023autogen, hong2024metagpt, qian2024chatdev, li2023camel, park2023generative}.
Recent work has turned to agentic loops and ensemble scaling:
\citet{yao2023react} interleave reasoning and acting,
\citet{shinn2023reflexion} add verbal self-critique,
\citet{zhang2024chainofagents} specialize the chain to long-context
reasoning, and \citet{li2024moreagents} show that sampling-and-voting
ensembles improve accuracy monotonically with agent count. Most
directly, \citet{cemri2025multiagent} catalog fourteen failure modes
of multi-agent LLM systems and argue that current evaluation cannot
anticipate them; our structural diagnostic offers one principled
approach to that gap. Across this literature, topology is treated as
a design parameter chosen by trial and error. We sit upstream of any
specific prompting strategy and ask what the communication graph
alone tells us about failure modes, before any inference is run.

\textbf{Evaluation of LLM systems.} HELM \citep{liang2023helm} and
BIG-Bench \citep{srivastava2023beyond} measure single-model
capability; AgentBench \citep{liu2025agentbench} and ChatEval
\citep{chan2023chateval} target multi-agent protocols. These
instruments are outcome-oriented and post hoc. We pursue a
pre-inference diagnostic derived from the graph itself, complementary
to outcome-based evaluation rather than a substitute for it.

\textbf{Spectral analysis of message-passing systems.} Spectral
approaches to information flow on graphs are well established in
graph signal processing and graph neural networks
\citep{defferrard2016chebnet, kipf2017semi}, where they motivate
convolutional filter design and expressivity analysis. The closest
mechanistic analogue of our condition-number finding is over-squashing,
in which geometric bottlenecks compress long-range information
\citep{alon2021bottleneck, topping2022oversquashing}. The complementary
phenomenon of oversmoothing has been characterized empirically
\citep{li2018deeperGCN} and theoretically as an exponential loss of
expressive power with depth \citep{oono2020graph}, and recent work
unifies the two through width, depth, and topology, identifying high
commute time as the structural signature of over-squashing
\citep{digiovanni2023oversquashing}. We borrow the analytic posture
of this literature but transport it to a setting in which the
operator is a communication graph among LLM agents rather than a GNN
layer: a high-condition-number predictive map signals that
perturbations to the driving signal can be amplified along
ill-conditioned directions of the influence operator.

\textbf{Consensus and opinion dynamics on networks.} A classical
literature studies how distributed estimates converge under linear
pooling. The DeGroot model \citep{degroot1974consensus} establishes
that iterated weighted averaging on a strongly connected graph
converges to a consensus determined by the dominant eigenvector of
the transition matrix; subsequent work characterizes convergence
rates via the spectral gap and develops randomized variants
\citep{olfatisaber2007consensus, boyd2006gossip}, and the
mixing-time machinery is the same we use here
\citep{levin2017markov, chung1997spectral}. The robust-aggregation
literature is closer still: \citet{xiao2007distributed} characterize least-mean-square deviation in noisy distributed averaging, while  \citet{blanchard2017krum} introduce Krum as a Byzantine-tolerant aggregation rule. The malicious-leaf bound in Appendix~\ref{sec:malicious-bound} is the spectral counterpart of those results, identifying the structural quantity that controls how much a single adversarial agent can degrade the conditioning of the influence operator. Our spectral-gap diagnostic transposes consensus theory to LLM communication graphs, with the additional observation that explicit aggregation operators introduce bottlenecks not captured by the raw graph spectrum.

\textbf{Successor representations and predictive maps.} The SR was
introduced by \citet{dayan1993sr} as a compact summary of expected
discounted future state occupancy, and has become central to the
study of predictive cognitive maps, with evidence that the
hippocampus represents an SR-like quantity
\citep{stachenfeld2017hippocampus} and that human reinforcement
learning uses SR-like predictive structure
\citep{momennejad2017successor, russek2017predictive}. Successor
features generalize the construction to function approximation
\citep{barreto2017successor}, eigenoption frameworks use the SR
spectrum to discover temporal abstractions
\citep{machado2017eigenoptions}, and \citet{gershman2018successor}
surveys the computational logic that motivates treating the SR as a
general-purpose predictive representation. We bring this object to
multi-agent LLM systems, where the role of state-transition operator
is played by the row-stochastic adjacency of the communication graph
rather than by environment dynamics.

\textbf{Adversarial robustness on graphs and language models.}
\citet{zugner2018adversarial} introduce adversarial attacks on graph neural networks via edge and feature manipulation, and \citet{dai2018adversarial} extend the attack surface to RL-driven structural perturbations. For LLM-specific threat models, \citet{perez2022redteam} establish a precedent for using language models to red-team other language models. The malicious-leaf analysis in Appendix~\ref{sec:malicious-bound} sits at the intersection of these threads: it asks how a single adversarial agent in a fixed communication graph can manipulate the spectral diagnostics that govern downstream reasoning.

\textbf{Reasoning degradation under length.} Prior work documents that LLM reasoning degrades with horizon
\citep{dziri2023faith, liu2023lost}. Our chain results are
consistent and suggest that sequentiality in the communication graph is a separate, composable source of drift that interacts with single-agent length effects.

\section{Method}
\subsection{From Communication Graphs to Predictive Maps}
A multi-agent LLM system is represented as a directed graph $G = (V, E)$ with $|V| = n$ agents. The weighted adjacency $A \in \mathbb{R}^{n \times n}$ is row-normalized into a stochastic transition matrix $P$, where $P_{ij} = A_{ij} / \sum_{k} A_{ik}$. The successor representation with discount $\gamma \in [0, 1)$ is
\begin{equation}
M = (I - \gamma P)^{-1} = \sum_{k=0}^{\infty} \gamma^{k} P^{k},
\end{equation}
which aggregates $k$-step influence pathways weighted by $\gamma^{k}$. We fix $\gamma = 0.9$ throughout. In the predictive-map reading, $M$ tells us how a perturbation introduced at one agent is expected to propagate across the system over discounted horizons. It is an operator on state perturbations rather than on full reasoning trajectories, and this is deliberate: we want the diagnostic to be cheap, analytic, and independent of the language modality.

\subsection{Three Spectral Readings of the Predictive Map}
From $M$ we extract three scalar summaries, each with a distinct mechanistic reading.

\textbf{Spectral radius $\rho(M) = \max_i |\lambda_i|$.} In the linear-systems view, $\rho(M)$ bounds the geometric growth rate of perturbations under repeated application. For a row-stochastic $P$ with $\rho(P) = 1$ and ergodic structure, $\rho(M) = (1 - \gamma)^{-1}$. Nilpotent operators, such as the acyclic chain, depart from this bound: $\rho(M_{\text{chain}}) = 1$. Read mechanistically, $\rho$ separates acyclic from cyclic communication and offers a first, cautious prediction of error-amplification tendency. We will see that this prediction is empirically inverted, and we develop a drift-corrected variant of $\rho$ in Section \ref{sec:affine}.

\textbf{Spectral gap $\Delta(M) = |\lambda_1| - |\lambda_2|$.} The gap governs the mixing time of the underlying chain and, by analogy, the speed at which distributed estimates converge to a shared state \citep{levin2017markov, chung1997spectral}. Read mechanistically, it predicts the dynamics of consensus.

\textbf{Condition number $\kappa(M) = \sigma_{\max}(M) / \sigma_{\min}(M)$.} $\kappa$ is the classical measure of sensitivity to perturbations of the driving signal. Read mechanistically, it predicts robustness under controlled perturbation.

\subsection{Empirical Counterparts}
For each spectral reading we define a task-level measurement, so that the predictions are falsifiable. Cumulative error growth is $E_{\mathrm{ceg}} = \sum_{t=1}^{T} |x_{t}^{\star} - \hat{x}_{t}|$, the sum of per-step deviations from the ground-truth state. Consensus decay is $R_{\mathrm{cdr}} = \frac{1}{T - 1} \sum_{t=1}^{T - 1} \log(D_{t+1} / D_{t})$, the mean log-rate at which average pairwise disagreement $D_{t}$ contracts over steps. Perturbation sensitivity is $F_{\mathrm{ps}} = |x_{T}' - x_{T}|$, the divergence of the final state under a controlled input perturbation. Full definitions are given in Appendix \ref{app:empirical-math}.

\subsection{What the Framework Predicts}
Our claim is local and precise. On the state-tracking task described in Section \ref{sec:setup}, and with the model and decoding parameters described there, the rank ordering of the three empirical measurements across the three topologies should agree with the rank ordering of the corresponding spectral quantity, up to any inversion that the predictive-map view itself makes visible. The spectral-radius case is where we expect such an inversion, and it is where we find one.

\subsection{An Affine-Noise Model and a Drift-Corrected Gain}
\label{sec:affine}
The three spectral readings above treat $M$ as an operator on perturbations of a deterministic flow. Sequential LLM reasoning departs from this picture in a way that is empirically large but spectrally invisible: each agent introduces its own stochastic deviation from the true rule $\tau$, and topologies differ in how those deviations compose. We model this explicitly. At step $t$, agent $i$ produces
\begin{equation}
\hat{x}_t^{(i)} = \tau(\hat{x}_{t-1}) + \eta_t^{(i)}, \qquad \eta_t^{(i)} \sim \mathcal{D}(0, \sigma^2),
\label{eq:affine}
\end{equation}
where $\mathcal{D}$ is a zero-mean noise distribution with variance $\sigma^2$, iid across agents and steps. The noise term aggregates per-agent arithmetic and decoding stochasticity; equation \eqref{eq:affine} is the simplest model that captures the ``small non-zero deviation per agent'' picture without committing to a particular error mechanism.

For a topology that aggregates $k$ agents per step into a single state via averaging, the effective per-step noise has variance $\sigma^2 / k$. For the chain, $k = 1$. Assuming $\tau$ is approximately Lipschitz with constant near unity over the operating range (which is consistent with the bounded state of our task), iterating equation \eqref{eq:affine} for $T$ steps gives
\begin{equation}
\mathbb{E}\!\left[ E_{\mathrm{ceg}} \right] \;\approx\; c \cdot \frac{\sigma}{\sqrt{k}} \cdot T^{3/2},
\label{eq:eceg-prediction}
\end{equation}
for a constant $c$ that depends on $\tau$ and $\mathcal{D}$ but not on the topology. We give the derivation in Appendix \ref{app:affine}. Equation \eqref{eq:eceg-prediction} predicts that, for fixed $T$ and $\sigma$, the cumulative error ratio between a chain and a topology with $k$-fold aggregation is $\sqrt{k}$. With $k = 4$ leaves per star step and $k = 4$ peer agents per mesh step, the predicted chain-to-aggregated ratio is $2$, which matches the empirical chain/star and chain/mesh ratios in Table \ref{tab:empirical} to within sampling variance.

We summarize this with a drift-corrected gain
\begin{equation}
\tilde{\rho}(M; \mathbf{k}) \;=\; \rho(M) \cdot \sqrt{ \tfrac{1}{n} \textstyle\sum_i 1/k_i },
\label{eq:tilde-rho}
\end{equation}
where $k_i$ is the effective aggregation count at agent $i$ (one for unaggregated nodes, the in-degree of the aggregator at aggregation steps). On the chain, the aggregation factor is $1$ throughout; on the star and mesh, the aggregation factor contracts $\tilde{\rho}$ by approximately $1/\sqrt{4} = 0.5$ relative to a chain of equivalent depth. The rank order of $\tilde{\rho}$ across the three topologies matches the empirical rank order of $E_{\mathrm{ceg}}$, recovering predictive validity that the bare spectral radius lacks. We use $\tilde{\rho}$ as a fourth, derived diagnostic alongside $\rho$, $\Delta$, and $\kappa$ in the results below.

Two extensions of this framework, to correlated per-agent noise and
to dynamically weighted transition operators, are developed in
Appendices~\ref{sec:correlated-noise} and~\ref{sec:dynamic-attention}
respectively.

\section{Experimental Setup}
\label{sec:setup}
\textbf{Task.} Each trial executes a 12-step update of a structured JSON state with three fields: a floating-point \texttt{Value}, a binary \texttt{Parity} $\in \{A, B\}$, and a bounded integer \texttt{Level} $\in [1, 9]$. A three-rule, order-dependent transition protocol couples arithmetic precision, conditional branching, and bounded state updates. The choice of a three-field state is motivated by findings in cognitive psychology that active working memory is limited to approximately four plus or minus one maintained elements \citep{cowan2001magical}; saturating this capacity at each step discourages shallow pattern completion. Agent contexts are reset between steps so that errors propagate only through the explicit communication channel.

\textbf{Topologies.} We evaluate three canonical structures. The \emph{chain} executes 12 sequential agents, one per step. The \emph{star} uses four leaf agents per step whose proposals a central judge aggregates \citep{chan2023chateval}. The \emph{mesh} adds a second deliberation pass in which agents critique peer proposals before a majority vote \citep{du2023debate}. Agent counts follow prior multi-agent deliberation work.

\textbf{Model and decoding.} All agents are local Qwen2.5-7B-Instruct \citep{qwen2} calls (commit \texttt{a09a354}) with temperature $0.8$, top-$p$ $0.5$, and a 250-token budget. Perturbation sensitivity is evaluated with $\varepsilon = 15.0$ applied to \texttt{Value}.

\textbf{Protocol.} Each (topology, condition) pair is executed for 100 independent trials. Reported means and standard deviations are over trials. The topology effect is assessed by a Kruskal--Wallis $H$ test, and the rank-order agreement between theory and experiment by Spearman $r_s$ over three ordered topology means; we treat $r_s$ here as rank-consistency evidence rather than as a significance test, since three ranks offer essentially no statistical power.

\textbf{Compute.} Experiments ran on a single NVIDIA A100 (32 GB). A full run (three topologies, 100 trials, 12 steps) completes within 6 wall-clock hours.

\begin{figure}[htbp]
\centering
\includegraphics[width=\linewidth]{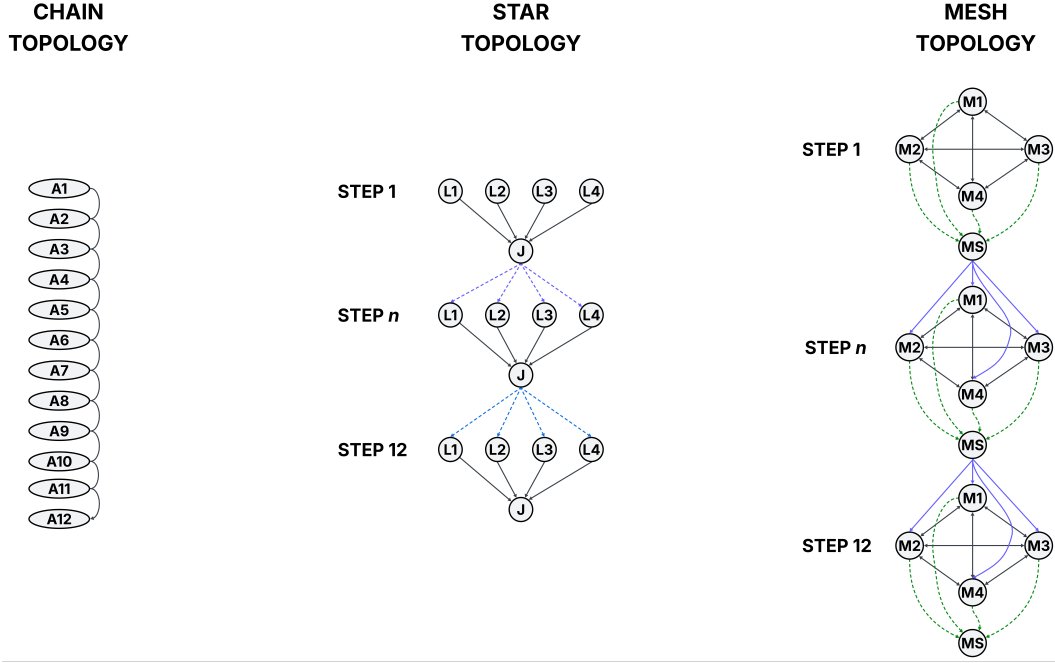}
\caption{Communication topologies used in the study: a 12-agent chain, a four-leaf star with judge aggregation, and a four-agent mesh with peer deliberation and majority vote.}
\label{fig:topologies}
\end{figure}

\section{Results}
\label{sec:results}
We first derive topology-specific predictions from the SR spectrum and then compare each prediction against the empirical distribution obtained over 100 trials.

\subsection{Spectral Predictions}
\label{subsec:spectral-predictions}
Closed-form derivations (Appendix \ref{app:derivations}) give the predictions in Table \ref{tab:spectral}. The chain's nilpotent adjacency collapses its spectral radius to 1 and its spectral gap to 0, while the star and mesh, both ergodic under row-stochastic normalization, saturate $\rho(M) = (1 - \gamma)^{-1} = 10$.

\begin{table}[htbp]
\centering
\caption{Spectral diagnostics of the successor representation for each topology, at $\gamma = 0.9$.}
\label{tab:spectral}
\begin{tabular}{lccc}
\toprule
\textbf{Topology} & $\rho(M)$ & $\Delta(M)$ & $\kappa(M)$ \\
\midrule
Chain & 1.00 & 0.00 & 9.95 \\
Star  & 10.00 & 9.00 & 28.61 \\
Mesh  & 10.00 & 9.23 & 13.00 \\
\bottomrule
\end{tabular}
\end{table}

\subsection{Empirical Measurements}
\label{subsec:empirical-measurements}
The empirical means over 100 trials, together with their rank-consistency statistics against the predictions in Table \ref{tab:spectral}, are collected in Table \ref{tab:empirical}. The topology effect is strongly significant for all three metrics.

\begin{table}[htbp]
\centering
\caption{Empirical measurements (mean $\pm$ s.d.\ over 100 trials) and rank-consistency statistics against the spectral predictions in Table \ref{tab:spectral}.}
\label{tab:empirical}
\small
\setlength{\tabcolsep}{3.5pt}
\begin{tabular}{p{3.0cm}cccccc}
\toprule
\textbf{Metric} & \textbf{Chain} & \textbf{Mesh} & \textbf{Star} & $H$ & $p$ & $r_s$ \\
\midrule

Error amplification \\ ($E_{\mathrm{ceg}}$)
& \makecell[c]{2094.34 \\ $\pm$ 843.05}
& \makecell[c]{1241.00 \\ $\pm$ 950.01}
& \makecell[c]{1184.24 \\ $\pm$ 951.91}
& 58.02 & $<0.001$ & $-1.00$ \\

Consensus decay \\ ($R_{\mathrm{cdr}}$)
& \makecell[c]{0.27 \\ $\pm$ 0.08}
& \makecell[c]{-1.66 \\ $\pm$ 42.25}
& \makecell[c]{-3.44 \\ $\pm$ 2.40}
& 188.68 & $<0.001$ & $+0.50$ \\

Perturbation sensitivity \\ ($F_{\mathrm{ps}}$)
& \makecell[c]{237.82 \\ $\pm$ 190.51}
& \makecell[c]{247.42 \\ $\pm$ 198.51}
& \makecell[c]{443.64 \\ $\pm$ 307.45}
& 29.03 & $<0.001$ & $+1.00$ \\

\bottomrule
\end{tabular}
\end{table}

\subsection{Spectral Radius and the Stability Paradox}
Read naively, the spectral radius predicts that the chain ($\rho = 1$) should be the most stable topology under repeated information flow. Empirically, the chain exhibits the largest cumulative error, roughly twice the mesh and star, and the rank correlation between prediction and observation is $-1$. The disagreement is not a numerical artifact; it is a consequence of which kind of stability the spectral radius encodes. The spectral radius governs the geometric growth of \emph{deterministic} perturbations to a homogeneous linear flow. The cumulative error in our task is dominated instead by accumulation of \emph{stochastic} per-agent deviations, which is invisible to the homogeneous spectrum.

The affine-noise model of Section \ref{sec:affine} makes this precise. Equation \eqref{eq:eceg-prediction} predicts $\mathbb{E}[E_{\mathrm{ceg}}] \propto \sigma T^{3/2} / \sqrt{k}$, where $k$ is the per-step aggregation count. For the chain ($k = 1$) versus star and mesh ($k = 4$), this gives a chain-to-aggregated ratio of $2$, which is what we observe. The drift-corrected gain $\tilde{\rho}$ defined in equation \eqref{eq:tilde-rho} rank-orders the three topologies as chain $>$ star $\approx$ mesh, in agreement with the empirical ranking of $E_{\mathrm{ceg}}$.

We read this as a structural finding rather than a failure of the framework: linear spectral stability and stochastic-drift stability are distinct notions, and a topology can be highly stable in the first sense while being maximally fragile in the second. Pipelines whose dominant error source is per-agent stochasticity, which we expect to be most LLM pipelines, are governed by drift stability rather than spectral stability. The result is a call for representational diagnostics, of which $\tilde{\rho}$ is one example, that sit alongside classical spectral ones.

\begin{figure}[H]
\centering
\includegraphics[width=0.75\linewidth]{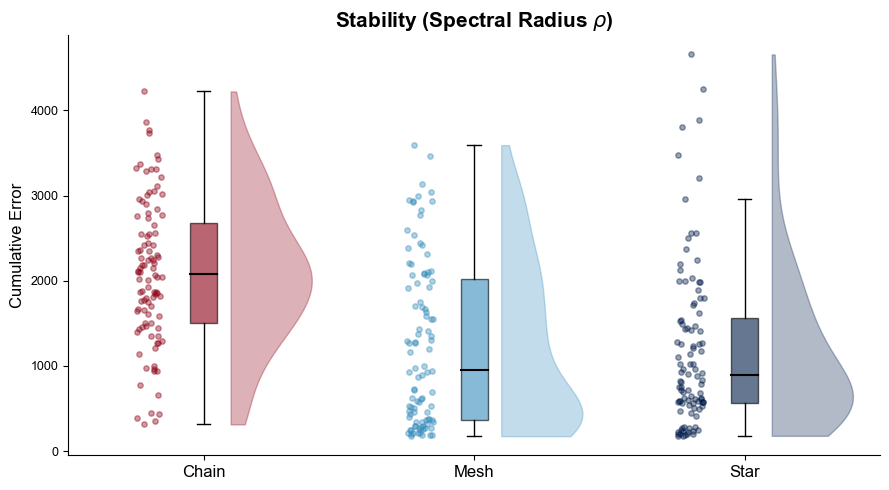}
\caption{Cumulative error growth across 100 trials. The chain shows the largest error despite the smallest spectral radius, illustrating the stability paradox.}
\label{fig:sr}
\end{figure}

\subsection{Spectral Gap and Consensus}
The spectral gap predicts fastest convergence for the mesh, then the star, then the chain. The empirical ordering places the chain last as predicted, but inverts the relative position of star and mesh. We read this through the same representational lens: the star's many-to-one judge aggregation imposes a geometric bottleneck that accelerates consensus beyond what the raw mesh gap alone would suggest. The gap captures mixing on the graph, not the effect of explicit aggregation operators inserted into the flow.

\begin{figure}[H]
\centering
\includegraphics[width=0.75\linewidth]{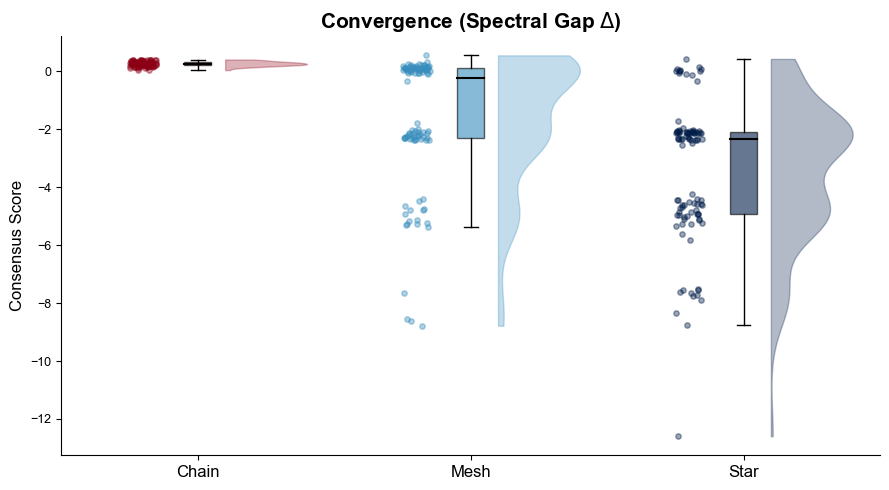}
\caption{Consensus decay rate by topology. The chain fails to reduce disagreement; star and mesh both converge, with the star's judge bottleneck accelerating aggregation.}
\label{fig:sg}
\end{figure}

\subsection{Condition Number and Perturbation Robustness}
The condition-number ordering $\kappa_{\text{chain}} < \kappa_{\text{mesh}} < \kappa_{\text{star}}$ matches the empirical perturbation-sensitivity ordering exactly. Within this case study, $\kappa$ is the cleanest diagnostic we observed, and it is the one we would put first in any practical pre-inference triage.

\begin{figure}[H]
\centering
\includegraphics[width=0.75\linewidth]{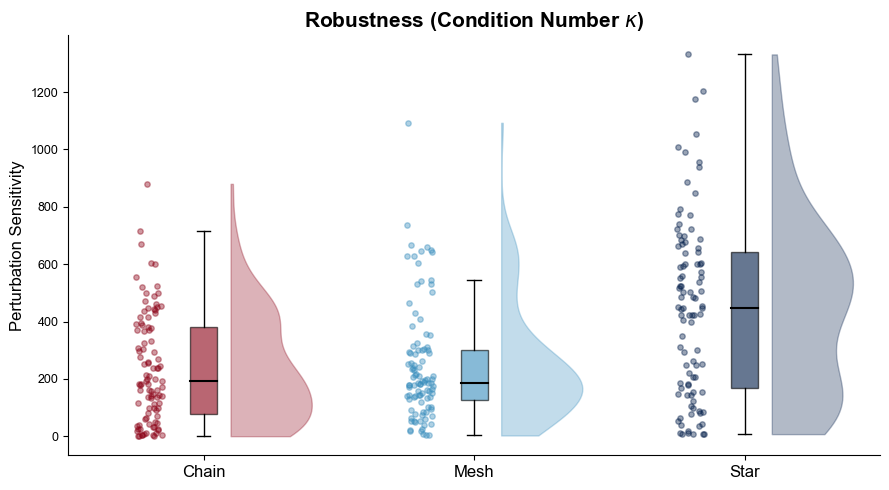}
\caption{Perturbation sensitivity across topologies. The empirical ordering matches the condition-number prediction exactly.}
\label{fig:cn}
\end{figure}

\subsection{A Practical Triage Rule}
The findings above suggest a simple pre-inference triage. Compute $\kappa(M)$ first: it was the cleanest predictor of perturbation sensitivity in our experiments, and a single scalar from a closed-form linear-algebraic operation suffices to rank candidate topologies on adversarial robustness. Compute $\Delta(M)$ next, with the caveat that explicit aggregation operators (judges, majority votes) introduce bottlenecks the raw gap does not see; for topologies that contain such operators, treat $\Delta$ as a lower bound on consensus speed. For cumulative error, do not rely on $\rho(M)$ alone; use $\tilde{\rho}(M; \mathbf{k})$ from equation \eqref{eq:tilde-rho}, which factors in the per-step aggregation count and is the diagnostic whose ranking agreed with our empirical $E_{\mathrm{ceg}}$. The full computation requires only the row-stochastic adjacency and the in-degree profile of aggregation nodes, and runs in milliseconds for graphs of practical size.

\section{Discussion}
The results admit a unified reading. The successor representation of a communication graph is a compact summary of how influence propagates, and three of its spectral quantities pick out three distinct aspects of that propagation. Two of them, the spectral gap and the condition number, behave as the linear intuition predicts and align with classical consensus and conditioning theory. The third, the spectral radius, is empirically inverted, and the inversion is informative rather than disqualifying: it locates the regime in which the homogeneous linear spectrum is blind to inhomogeneous stochastic drift. The affine-noise model of Section \ref{sec:affine} explains the inversion quantitatively and yields a corrected gain $\tilde{\rho}$ whose ranking matches observation.

Three lines of extension are natural. The first is to sweep $\gamma$, the agent count $n$, and the aggregation count $k$ to test the $\sqrt{k}$ prediction in equation \eqref{eq:eceg-prediction} directly, rather than only at the three points implied by the chain, star, and mesh. The second is to enrich the noise model: per-agent biases that are correlated across agents (for instance, because all agents share a base model with a shared inductive error) violate the iid assumption and should change the aggregation gain, and our framework offers a controlled setting in which to study this. The third is to move from row-stochastic to weighted graphs in which edge weights encode trust, attention, or estimated agent reliability, which would let the operator $P$ itself become a learned object and bring the diagnostic closer to the design loop in which topologies are chosen.

The three spectral quantities can also be read as structural
counterparts of three recurring concerns in the trustworthy
multi-agent AI literature. Reliability under stochastic per-agent
deviation, the dominant failure mode in pipelines built from
homogeneous base models \citep{du2023debate, cemri2025multiagent},
is the regime that the drift-corrected gain $\tilde{\rho}$
addresses. Self-consistency across reasoning paths, the property
that debate and ensemble protocols are explicitly designed to
enforce \citep{chan2023chateval, li2024moreagents}, is the regime
that the spectral gap $\Delta$ tracks, with the caveat that
explicit aggregators induce bottlenecks the raw gap does not see.
Robustness to adversarial inputs and to inter-agent manipulation,
the concern that motivates red-teaming and Byzantine-resilient
aggregation \citep{perez2022redteam, blanchard2017krum}, is the regime in which the condition number $\kappa$ and the malicious-leaf bound of Appendix~\ref{sec:malicious-bound} are the relevant quantities. As LLM-based systems migrate from monolithic
models to multi-agent architectures, the relevant failure modes become structural properties of the communication graph rather than purely behavioral properties of any single agent
\citep{cemri2025multiagent}; pre-inference spectral diagnostics are one instantiation of that structural turn.

A more speculative reading connects this work to the cognitive-map literature from which the SR is drawn. If the SR captures a predictive map of an environment, and if the communication graph is the relevant ``environment'' for an ensemble of LLM agents, then the diagnostic we propose is a coarse approximation of the predictive map that the ensemble \emph{would} need to maintain in order to anticipate its own failure modes. We do not claim that current LLMs maintain such a map. We do suggest that giving the system designer access to one is a practically useful proxy.

\section{Limitations}
\label{sec:limitations}
We have deliberately scoped this study to a controlled regime in order to make the spectral predictions falsifiable. The cost of that scope is real and we state it plainly. The empirical study covers three topologies, one model family, one task family, and 100 trials per condition. A Spearman coefficient over three ranks has essentially no statistical power, and we report it only as rank-consistency evidence. Our empirical metrics reduce full reasoning trajectories to scalars and discard information that a richer trajectory-level analysis would retain. The perturbation-sensitivity effect ranks as predicted. Mesh deliberation parameters follow prior work but are not extensively swept. The affine-noise model assumes iid noise across agents and a Lipschitz-near-unity transition rule; both are simplifications that may break for tasks with strong nonlinearity or for systems in which agents share systematic biases. We expect the qualitative ordering to be model-agnostic where the dominant error source is per-agent stochasticity, since the affine-noise derivation depends on $P$ and on noise variance rather than on any property of the underlying language model, but this expectation is a prediction of the framework rather than an empirical finding and we flag it as the most important sweep for follow-up work. Claims about model scale, closed-weight systems, or topologies outside the chain, star, and mesh triple are not supported by the present evidence and we identify the corresponding sweeps as the natural next step.

\section{Broader Impact}
\label{sec:broader-impact}
Structural pre-inference diagnostics can help practitioners choose safer multi-agent configurations, reducing wasted compute and the risk of deploying pipelines whose failure modes are discovered only post hoc. The same tools could in principle be used to optimize adversarial multi-agent systems for maximal perturbation amplification; we note the possibility and recommend that downstream work pair structural diagnostics with explicit evaluation of adversarial robustness. Our experiments use only open-weight models and synthetic task data, and involve no human subjects or personally identifying information.

\section{Reproducibility}
We release all configuration files and evaluation harnesses, and we pin the exact model commit (\texttt{Qwen/Qwen2.5-7B-Instruct}@\texttt{a09a354}). Appendix \ref{app:repro} lists hardware, software versions, decoding parameters, and the single command required to reproduce Tables \ref{tab:spectral} and \ref{tab:empirical}.

\section{Conclusion}
We have argued that a predictive map of the communication graph, constructed as a successor representation and summarized by three spectral quantities, is a useful pre-inference diagnostic for multi-agent LLM systems. In a controlled case study across chain, star, and mesh topologies, the condition number rank-predicts empirical perturbation robustness perfectly, the spectral gap partially tracks consensus, and the spectral radius is inverted with respect to cumulative error. The inversion marks a stability paradox that points beyond linear spectra toward representational, drift-aware diagnostics; we proposed a drift-corrected gain $\tilde{\rho}$ derived from an affine-noise model that recovers the empirical ordering. Establishing the generality of this picture, across models, tasks, and richer topologies, is the natural next step.

\newpage
\bibliographystyle{unsrtnat}
\bibliography{references}

\newpage
\appendix
\section{Theoretical Derivations}
\label{app:theory}

This appendix collects the mathematical support for the main-text claims. Section \ref{app:derivations} derives the closed-form spectra reported in Table \ref{tab:spectral}. Section \ref{app:empirical-math} states the empirical metrics defined in Section \ref{sec:setup}. Section \ref{app:affine} derives the affine-noise prediction of Section \ref{sec:affine}.

\subsection{Closed-Form Spectra of the Three Topologies}
\label{app:derivations}

\paragraph{Chain.} $A_{\text{chain}}$ is the $12 \times 12$ upper shift matrix. Its row-normalized form $P_{\text{chain}}$ is nilpotent of index 12, so $P^{k} = 0$ for $k \geq 12$ and $M_{\text{chain}} = \sum_{k=0}^{11} \gamma^{k} P^{k}$ is upper triangular with diagonal entries 1 and superdiagonal entries $\gamma^{k}$. Hence $\rho(M_{\text{chain}}) = 1$ and $\Delta(M_{\text{chain}}) = 0$; numerical evaluation at $\gamma = 0.9$ gives $\kappa(M_{\text{chain}}) \approx 9.95$.

\paragraph{Star.} For a star with one center and $\ell$ leaves, the row-stochastic transition matrix has eigenvalues $\{1, 0, \dots, 0, -1\}$ with $\ell - 1$ zeros. The SR eigenvalues are $\{(1 - \gamma)^{-1}, 1, \dots, 1, (1 + \gamma)^{-1}\}$, giving $\rho = 10$, $\Delta = 9$, and $\kappa \approx 28.61$ in the experimental configuration.

\paragraph{Mesh.} The complete graph on $n = 4$ with uniform off-diagonal weights has eigenvalues $\{1, -\tfrac{1}{3}, -\tfrac{1}{3}, -\tfrac{1}{3}\}$. The SR then has eigenvalues approximately $\{10, 0.769, 0.769, 0.769\}$, giving $\rho = 10$, $\Delta \approx 9.23$, $\kappa \approx 13.00$.

\subsection{Empirical Metric Definitions}
\label{app:empirical-math}

\paragraph{Cumulative error growth.} With $\epsilon_{t} = |x_{t}^{\star} - \hat{x}_{t}|$, $E_{\mathrm{ceg}} = \sum_{t=1}^{T} \epsilon_{t}$.

\paragraph{Consensus decay.} For $N$ agents, $D_{t} = \frac{1}{N(N-1)} \sum_{i \neq j} |s_{i,t} - s_{j,t}|$, $r_{t} = \log(D_{t+1} / D_{t})$, $R_{\mathrm{cdr}} = \frac{1}{T - 1} \sum_{t} r_{t}$. For the chain, $D_{t} = |x_{t}^{\star} - \hat{x}_{t}|$.

\paragraph{Perturbation sensitivity.} $F_{\mathrm{ps}} = |x_{T}' - x_{T}|$, where $x_{T}'$ is the final value under a controlled input perturbation of magnitude $\varepsilon = 15.0$ applied to \texttt{Value}.

\subsection{Affine-Noise Derivation}
\label{app:affine}

We derive equation \eqref{eq:eceg-prediction} from the affine-noise model of Section \ref{sec:affine}. Let $\tau: \mathbb{R} \to \mathbb{R}$ denote the deterministic transition rule, and assume $\tau$ is Lipschitz with constant $L \leq 1$ on the operating range of the state, so that errors do not amplify under iteration. Let $\hat{x}_t$ denote the state estimate at step $t$ and $x_t^\star = \tau^t(x_0)$ the ground-truth state. Define $e_t = \hat{x}_t - x_t^\star$.

\paragraph{Chain.} A single agent updates the state at each step:
\[
\hat{x}_t = \tau(\hat{x}_{t-1}) + \eta_t, \qquad \eta_t \sim \mathcal{D}(0, \sigma^2).
\]
Linearizing $\tau$ about $x_{t-1}^\star$ with local slope at most $L$, we have $e_t = L \cdot e_{t-1} + \eta_t$, hence
\[
\mathrm{Var}(e_t) \leq \sigma^2 \sum_{s=0}^{t-1} L^{2s} \leq \sigma^2 \cdot t \quad \text{for } L \leq 1.
\]
Then $\mathbb{E}|e_t| \leq C \sigma \sqrt{t}$ for a constant $C$ depending on $\mathcal{D}$, and
\[
\mathbb{E}\!\left[ E_{\mathrm{ceg}}^{\mathrm{chain}} \right] = \sum_{t=1}^T \mathbb{E}|e_t| \leq C \sigma \sum_{t=1}^T \sqrt{t} \leq \tfrac{2}{3} C \sigma\, T^{3/2}.
\]

\paragraph{Aggregated topology.} A topology with $k$-fold aggregation per step replaces the per-step noise $\eta_t$ with the average $\bar{\eta}_t = \frac{1}{k}\sum_{i=1}^k \eta_t^{(i)}$, which has variance $\sigma^2 / k$ under iid noise. The same derivation as above gives
\[
\mathbb{E}\!\left[ E_{\mathrm{ceg}}^{\mathrm{agg}} \right] \leq \tfrac{2}{3} C \cdot \frac{\sigma}{\sqrt{k}}\, T^{3/2}.
\]

\paragraph{Ratio.} Taking the ratio of the chain and aggregated bounds,
\[
\frac{\mathbb{E} E_{\mathrm{ceg}}^{\mathrm{chain}}}{\mathbb{E} E_{\mathrm{ceg}}^{\mathrm{agg}}} = \sqrt{k},
\]
which for $k = 4$ predicts a factor of $2$. The empirical chain-to-star and chain-to-mesh ratios in Table \ref{tab:empirical} are $1.77$ and $1.69$ respectively, both within sampling variance of the predicted $2$.

The derivation makes explicit which assumptions matter. The Lipschitz-near-unity assumption ensures errors do not amplify intrinsically; the iid noise assumption ensures that aggregation cleanly reduces variance; and the linearization is valid only when individual errors are small relative to the curvature of $\tau$. Violations of any of these (for instance, agents with correlated systematic biases, or rules with sharp branching) would change the prediction in directions that the framework can be extended to capture.

\subsection{Correlated Noise and a Bias-Aware Drift-Corrected Gain}
\label{sec:correlated-noise}

The iid assumption in Section 3.5 is suspect when agents share a  base model. \citet{du2023debate} show that ensembles of identical  LLMs converge to a common posterior reflecting shared inductive
priors rather than ground truth, so the per-agent error decomposes  naturally into a shared and an independent component. We model this
by writing $\eta_t^{(i)} = b_t + \xi_t^{(i)}$ with $b_t \sim
\mathcal{D}_b(0, \sigma_b^2)$ shared across agents at step $t$ and
$\xi_t^{(i)} \sim \mathcal{D}_\xi(0, \sigma_\xi^2)$ iid, giving total
variance $\sigma^2 = \sigma_b^2 + \sigma_\xi^2$ and inter-agent
correlation $\rho_c = \sigma_b^2 / \sigma^2 \in [0, 1]$.

For a $k$-fold aggregator, the aggregated noise variance is
\begin{equation}
\operatorname{Var}(\bar\eta_t)
= \frac{\sigma^2}{k}\bigl[1 + (k-1)\rho_c\bigr],
\label{eq:agg-variance-correlated}
\end{equation}
which interpolates between $\sigma^2 / k$ at $\rho_c = 0$ and
$\sigma^2$ at $\rho_c = 1$. The chain-to-aggregated cumulative-error
ratio of Appendix~A.3 generalizes from $\sqrt{k}$ to
$\sqrt{k / [1 + (k-1)\rho_c]}$, which collapses to unity as
$\rho_c \to 1$. With $k=4$ and the empirical chain-to-star ratio of
$1.77$, the inferred upper bound is $\rho_c \lesssim 0.06$. The
drift-corrected gain becomes
\begin{equation}
\tilde\rho_c\bigl(M;\, \{k_i\},\, \rho_c\bigr)
= \rho(M) \cdot
\sqrt{\frac{1}{n} \sum_{i=1}^{n}
\frac{1 + (k_i - 1)\rho_c}{k_i}},
\label{eq:rho-tilde-correlated}
\end{equation}
which recovers $\tilde\rho(M; k)$ at $\rho_c = 0$ and $\rho(M)$ at
$\rho_c = 1$, in which limit aggregation is structurally inert and
the original spectral-radius prediction is restored exactly.

\subsection{Dynamic Graph Attention and Reliability-Weighted Transitions}
\label{sec:dynamic-attention}

Static $P$ encodes only graph structure, not running deliberation
state. Following \citet{velickovic2018graph}, we lift $P$ to a
time-varying operator by parameterizing the edge from agent $i$ to
agent $j$ at step $t$ via
\begin{equation}
e_{ij}^{(t)}
= \operatorname{LeakyReLU}\!\left(
\mathbf{a}^\top \bigl[ \mathbf{W} \mathbf{h}_i^{(t)} \,\Vert\,
\mathbf{W} \mathbf{h}_j^{(t)} \bigr]
\right)
+ \beta\, \phi\bigl(r_j^{(t)}\bigr),
\quad
P_{ij}^{(t)}
= \frac{\exp\bigl(e_{ij}^{(t)}\bigr)}
{\sum_{k \in \mathcal{N}(i)} \exp\bigl(e_{ik}^{(t)}\bigr)},
\label{eq:dynamic-P}
\end{equation}
where $\mathbf{h}_i^{(t)}$ is an agent feature vector, $r_i^{(t)}$ a
running reliability score, $\phi$ monotone, and $\beta \geq 0$
controls the reliability contribution. Each $P^{(t)}$ remains
row-stochastic, so the spectral diagnostics apply pointwise. The
successor representation generalizes to $M^{(t)} = I + \gamma
P^{(t)} M^{(t+1)}$, and three reductions are useful in practice: the
instantaneous $\kappa(M^{(t)})$, the time-averaged surrogate $\bar M
= (I - \gamma \bar P)^{-1}$, and the worst-case $\sup_t
\kappa(M^{(t)})$.

The construction admits a variance-reduction reading that links it
to Appendix~\ref{sec:correlated-noise}. If $r_j^{(t)}$ asymptotically  estimates $1/\sigma_j^2$, the aggregator variance $\sigma_b^2 +
\sum_j (\alpha_j^{(t)})^2 \sigma_j^2$ is minimized at $\alpha_j^\star
\propto 1/\sigma_j^2$, which the GAT softmax implements when
$\beta\,\phi(r_j) = -\log \sigma_j^2$. Reliability weighting therefore
suppresses idiosyncratic variance optimally but leaves the systemic
floor $\rho_c$ untouched, in agreement with \citet{du2023debate}.

\subsection{Spectral Bounds Under a Malicious Star Leaf}
\label{sec:malicious-bound}

Consider the star with one center and $\ell$ leaves, and suppose one leaf is adversarial and able to inflate its weight in the center's aggregation: $P_{01} = \alpha/(\alpha + \ell - 1)$ and
$P_{0j} = 1/(\alpha + \ell - 1)$ for $j \in \{2, \ldots, \ell\}$,
with $\alpha \geq 1$. The eigenvalues of $P$ remain
$\{1, 0, \ldots, 0, -1\}$ for any $\alpha > 0$, so $\rho(M)$ and
$\Delta(M)$ are insensitive to the attack; only $\kappa(M)$
responds, since $P$ is non-symmetric. Let $\mu^2 = (\alpha^2 +
\ell - 1)/(\alpha + \ell - 1)^2 - 1/\ell$ measure the deviation of the leaf weights from uniformity, with geometric ceiling
$\mu_{\max}^2 = 1 - 1/\ell$ as $\alpha \to \infty$. A trace-bound argument on the active subspace of $A := I - \gamma P$ gives
\begin{equation}
\kappa(M) \;\leq\;
\frac{\bigl[\,3 + \gamma^2(\ell + 1/\ell + \mu^2)\,\bigr]^{3/2}}
{1 - \gamma^2},
\label{eq:kappa-bound}
\end{equation}
which at the worst-case $\mu^2 \to 1 - 1/\ell$ evaluates for
$\gamma = 0.9$, $\ell = 4$ to approximately $98.5$, a factor of
$3.4$ above the benign-star value $\kappa \approx 28.6$ from
Table~\ref{tab:spectral}. The bound scales as $\Theta(\ell^{3/2})$,
so larger stars are disproportionately vulnerable to weight
manipulation. Capping individual agent influence at $\bar w =
W/\ell$ enforces $\mu = 0$ structurally and recovers the benign
condition number as a hard upper bound.
\newpage
\section{Stepwise Reasoning Dynamics}
\label{app:stprd}

While aggregate metrics capture final system behavior, they do not fully reveal how reasoning trajectories evolve over intermediate steps. To better understand the dynamics of error propagation and consensus formation, we analyze stepwise traces of the empirical metrics across all trials.

For each topology, we plot the per-step metric values across all trials, along with the median trajectory to provide a robust summary of the central trend. We use the median rather than the mean to reduce sensitivity to outlier trajectories, which can arise from stochastic reasoning failures in individual runs.

\begin{figure}[H]
\vspace{-0.5em}
\centering
    \includegraphics[width=0.75\linewidth]{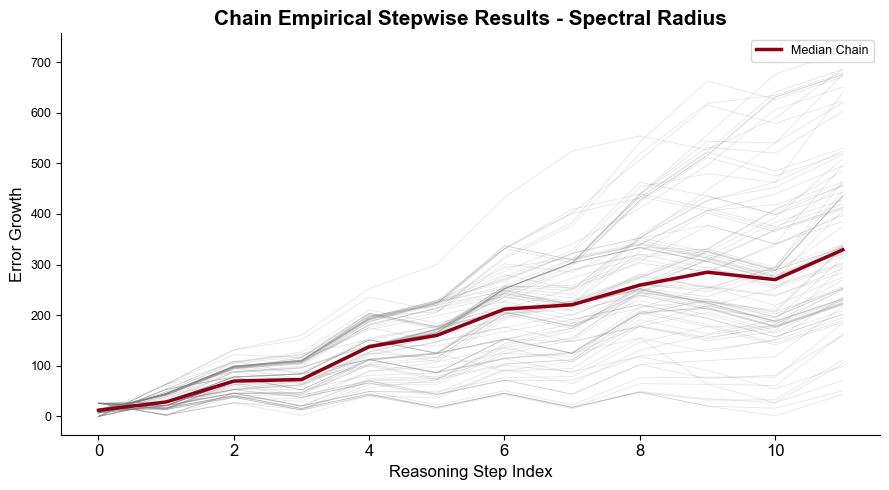}\\[-2.5ex]
    \includegraphics[width=0.75\linewidth]{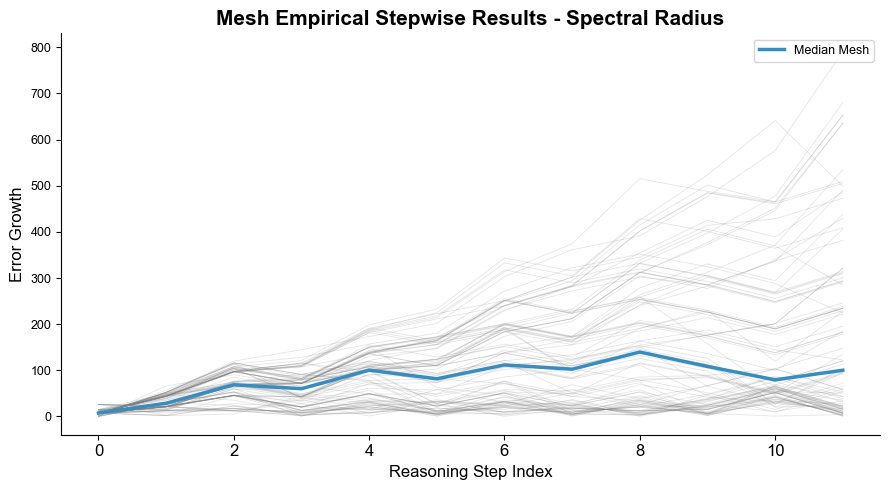}\\[-2.5ex]
    \includegraphics[width=0.75\linewidth]{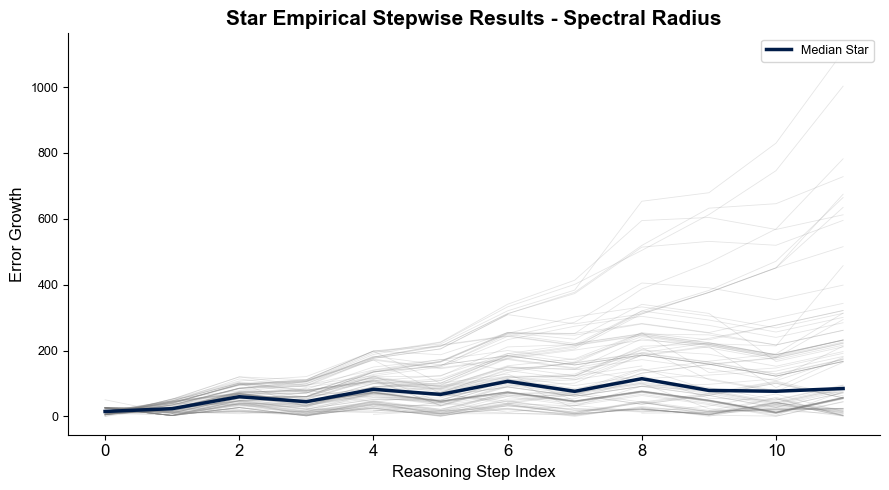}

    \vspace{-1em}
    \caption{Stepwise cumulative error growth for chain, mesh, and star topologies.}
    \label{fig:srstp_combined}
    \vspace{-0.5em}
\end{figure}

Figures \ref{fig:srstp_combined} (top and middle) and (bottom) show the evolution of cumulative error. The chain topology exhibits a steady monotonic increase, consistent with sequential error accumulation, while the star and mesh topologies display flatter trajectories.

The following figures illustrate the stepwise consensus dynamics. The star topology rapidly reduces disagreement early in the reasoning process, while the chain topology shows slower convergence, consistent with its zero spectral gap.

\begin{figure}[H]
\centering
    \includegraphics[width=0.75\linewidth]{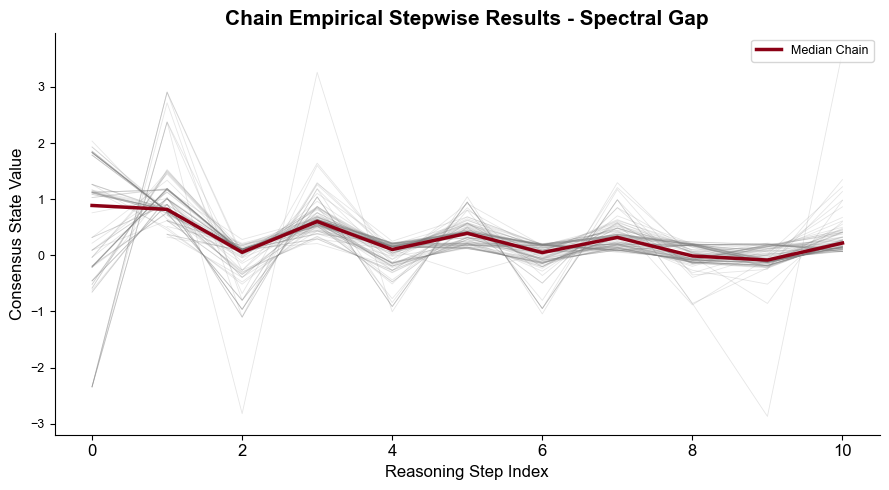}\\[-2.5ex]
    \includegraphics[width=0.75\linewidth]{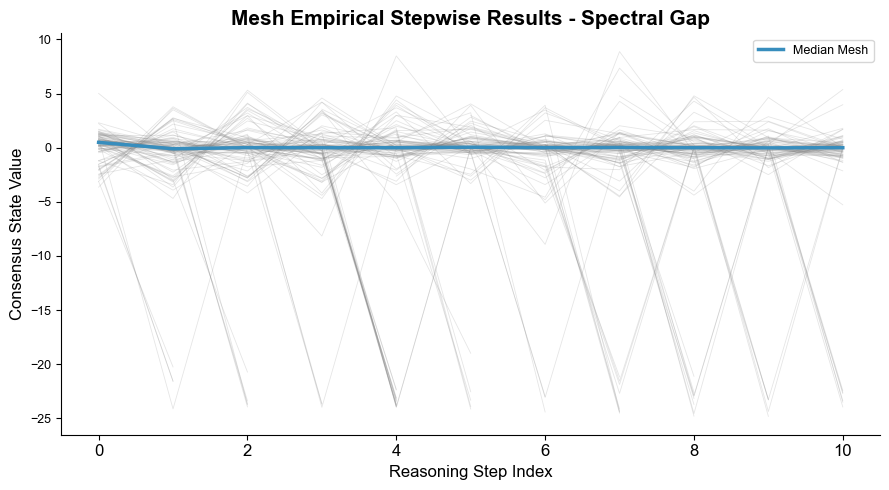}\\[-2.5ex]
    \includegraphics[width=0.75\linewidth]{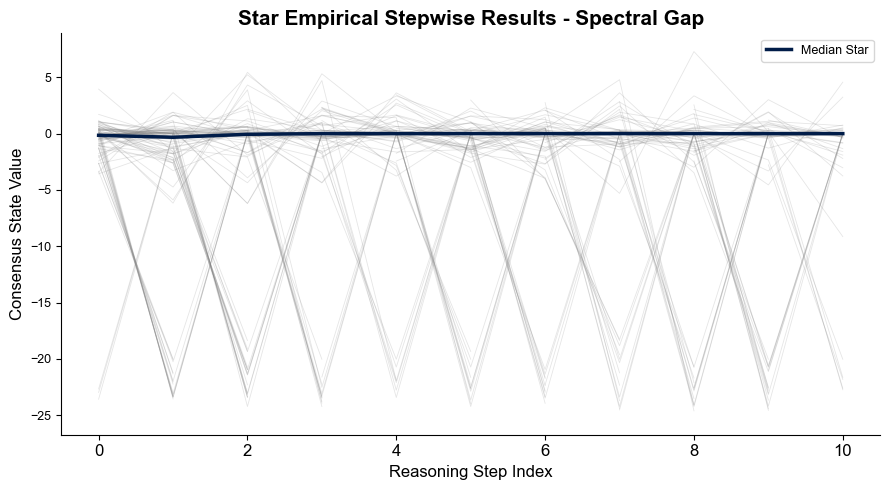}

    \caption{Stepwise consensus dynamics for chain (top), mesh (middle), and star (bottom) topologies.}
    \label{fig:sgstp_combined}
\end{figure}

\newpage
\section{State-Tracking Task Specification}
\label{app:sttsk}

A structured state-tracking task was designed to evaluate multi-step reasoning under controlled information flow. The task requires agents to iteratively update a shared JSON state across a fixed sequence of reasoning steps, where each update depends on the result of the previous step.

The state consists of three variables:
\begin{itemize}
    \item \textbf{Value}: a floating-point number requiring precise arithmetic updates
    \item \textbf{Parity}: a categorical variable taking values \{A, B\}
    \item \textbf{Level}: an integer constrained to a bounded range.
\end{itemize}

To ensure that reasoning errors propagate only through the communication topology, each agent receives only the current state and the fixed ruleset, without access to prior reasoning history beyond what is explicitly passed forward.

The base prompt used for each agent is provided below.

\begin{lstlisting}
[STRICT INSTRUCTION]
You are a JSON-only processor. Use the 'Reasoning' field to
calculate the new state before providing the final values.

STATE SCHEMA:
{
  "Reasoning": "Step-by-step math",
  "Value": <float>,
  "Parity": "<A or B>",
  "Level": <integer>
}

RULESET - apply in this exact order. Rules are dependent on each other.
Do not skip ahead. Each rule requires the result of the previous.

RULE 1 - Compute V_raw (do NOT round yet):
If Parity is "A": V_raw = (Value * 1.25) + (Level * 2)
If Parity is "B": V_raw = (Value * 0.75) - (Level * 2)
Do not round V_raw. Carry the full decimal forward into Rule 2.

RULE 2 - Compute V_new and P_new using V_raw from Rule 1:
Take the decimal portion of V_raw only (i.e. V_raw minus its integer part).
If that decimal portion is >= 0.5:
    V_new = round(V_raw + 1.5, 2)
If that decimal portion is < 0.5 but > 0.0:
    V_new = round(V_raw - 0.5, 2)
If the decimal portion is exactly 0.0:
    V_new = round(V_raw * 1.1, 2)
Then compute P_new:
If V_new < 70:  P_new = "A"
If V_new >= 70: P_new = "B"

RULE 3 - Compute L_new using BOTH V_new from Rule 2 AND P_new from Rule 2:
Apply EXCEPT logic carefully:
Apply L_new = Level + 2 (max 9) in ALL of the following cases EXCEPT
when P_new is "B" and V_new is between 60 and 90 inclusive,
in which case apply L_new = Level - 1 (min 1) instead.
The remaining cases are:
If Level >= 4 and V_new < 60: L_new = Level - 1  (minimum value is 1)
If Level < 4  and V_new < 60: L_new = Level - 2  (minimum value is 1)

FINAL OUTPUT FORMAT:
- {"Reasoning": "step-by-step math", "Value": <float>, "Parity": "<A or B>", "Level": <integer>}
\end{lstlisting}

In the star topology, multiple leaf agents independently process the same input state at each step. Their outputs are then passed to a central judge agent, which aggregates the proposals and produces the final state for the next step. The judge is allowed to perform its own reasoning and is not required to select one of the provided proposals.

\begin{lstlisting}
You are a judge agent responsible for determining the single correct JSON state.

You have received proposals from {n_leaf_agents} independent agents who each applied
the same ruleset to the same input state.

RULESET TO FOLLOW:
{rules_text}

PROPOSALS FROM LEAF AGENTS:
{json.dumps(proposals, indent=2)}

CURRENT INPUT STATE (what all leaf agents received):
{json.dumps(judge_state, indent=2)}

Review the proposals critically and calculate your own response if you feel there are errors in the responses.
Respond with ONLY a valid JSON object. No explanation, no extra text.
Format: {{"Value": <float>, "Parity": "<A or B>", "Level": <integer>}}
\end{lstlisting}

In the mesh topology, reasoning proceeds in two stages. In the first stage, all agents independently generate candidate outputs. In the second stage, each agent observes the proposals of all other agents and produces a revised output. The final state is determined through a majority-based aggregation of these second-round responses.

\begin{lstlisting}
You are a reasoning agent in a peer deliberation round.

In the first round, you and {n_agents - 1} other agents independently applied the same
ruleset to the same input state.

RULESET TO FOLLOW:
{rules_text}

INPUT STATE (what all agents received in round one):
{json.dumps(j_state, indent=2)}

FIRST-ROUND PROPOSALS FROM ALL AGENTS:
{json.dumps(first_pass, indent=2)}

Review the proposals critically and calculate your own response if you feel there are errors in the responses.
Respond with ONLY a valid JSON object. No explanation, no extra text.
Format: {{"Value": <float>, "Parity": "<A or B>", "Level": <integer>}}
\end{lstlisting}

\section{Reproducibility Details}
\label{app:repro}

All experiments ran on a single NVIDIA A100 (32 GB), Python 3.11.4, with dependencies pinned in \texttt{requirements.txt}. The full experimental suite is reproduced by:

\begin{lstlisting}
python rnnr_main.py --topology all \
  --mid_model Qwen/Qwen2.5-7B-Instruct \
  --task_difficulty enhanced --temp 0.8 --top_p 0.5 \
  --epsilon 15.0 --device cuda --n_trials 100 --n_workers 1 \
  --enforce_eager --max_tokens 250 \
  --out qwen25_full_spectral_run_t08_p05_e15_tr100.json
\end{lstlisting}

Minor variation across runs is expected due to decoding stochasticity; reported statistics are robust to this noise within the 100-trial bootstrap.

\end{document}